\RequirePackage{lineno} 
\documentclass{nature}
\usepackage{xcolor}
\usepackage{caption}
\usepackage{multirow}
\usepackage{amsmath}
\usepackage{CJK}
\usepackage{bm}
\usepackage{braket}
\usepackage{float}
\usepackage{amssymb}
\usepackage{color}
\usepackage{array}
\usepackage{ulem}
\usepackage{siunitx}
\usepackage{booktabs}
\usepackage{caption}
\usepackage{placeins}

\newcommand{\mrm}{\mathrm}
\newcommand{\mfk}{\mathfrak}

\def\ii{\mathrm{i}}

\def\kk{\mathbf{k}}

\DeclareSIUnit\angstrom{\text{Å}}

\title{Field-induced asymmetric band flattening and ideal quantum geometry in rhombohedral graphene}

\author{Hongyun Zhang$^{1,2,3,\dagger}$, Jinxi Lu$^{1,\dagger}$, Size Wu$^{4,\dagger}$, Yijie Wang$^{5,\dagger}$, Kai Liu$^{4}$, Fei Wang$^1$, Wanying Chen$^1$, Lingzhi Wen$^{1}$, Jinling Zhou$^{1}$, Kenji Watanabe$^6$, Takashi Taniguchi$^7$, Jose Avila$^{8}$, Pavel Dudin$^{8}$, Matthew D. Watson$^{9}$, Takafumi Sato$^{3,10}$, Pu Yu$^{1,11}$, Wenhui Duan$^{1,11,12}$, Zhida Song$^{5,*}$, Guorui Chen$^{4,*}$ \& Shuyun Zhou$^{1,11,*}$}

\usepackage{graphicx}
\makeatletter
\let\saved@includegraphics\includegraphics
\AtBeginDocument{\let\includegraphics\saved@includegraphics}

\makeatother

\begin{document}
\maketitle

\begin{affiliations}
    \item State Key Laboratory of Low-Dimensional Quantum Physics and Department of Physics, Tsinghua University, Beijing, People’s Republic of China
     \item Beijing Tsinghua Institute for Frontier Interdisciplinary Innovation, Beijing, People’s Republic of China
    \item Advanced Institute for Materials Research (WPI-AIMR), Tohoku University, Sendai, Japan
    \item State Key Laboratory of Micro-nano Engineering Science, Key Laboratory of Artificial Structures and Quantum Control (Ministry of Education), Tsung-Dao Lee Institute and School of Physics and Astronomy, Shanghai Jiao Tong University, Shanghai, People’s Republic of China
    \item International Center for Quantum Materials, School of Physics, Peking University, Beijing, People’s Republic of China
    \item Research Center for Electronic and Optical Materials, National Institute for Materials Science, Namiki, Tsukuba, Japan
    \item Research Center for Materials Nanoarchitectonics, National Institute for Materials Science, Namiki, Tsukuba, Japan
    \item Synchrotron SOLEIL, L’Orme des Merisiers, Saint Aubin-BP, Gif sur Yvette Cedex, France
	\item Diamond Light Source Ltd, Harwell Science and Innovation Campus, Didcot, UK
	\item Department of Physics, Graduate School of Science, Tohoku University, Sendai, Japan
	\item Frontier Science Center for Quantum Information, Beijing, People’s Republic of China
    \item Institute for Advanced Study, Tsinghua University, Beijing, People’s Republic of China\\
    $\dagger$ These authors contributed equally to this work.\\
    *Correspondence should be sent to syzhou@mail.tsinghua.edu.cn \& chenguorui@sjtu.edu.cn \& songzd@pku.edu.cn
\end{affiliations}

\begin{abstract}
Rhombohedral graphene exhibits an exceptionally diverse array of correlated phases that depend sensitively on the displacement field. Compiling reported phases into a unified phase diagram reveals a pronounced field-dependent electron-hole asymmetry: correlated states on the hole-doped side emerge at small displacement fields, whereas the fractional quantum anomalous Hall effect (FQAHE)\cite{Julong_Nature2024} is observed exclusively on the electron-doped side under large displacement fields. This stark asymmetry highlights the need to understand how flat bands evolve with displacement fields. 
Here, we directly visualize the field-induced electron-hole asymmetric band flattening in rhombohedral pentalayer graphene (R5G) using nanospot angle-resolved photoemission spectroscopy with electrostatic gating. Beyond gap opening and spectral weight redistribution indicative of layer polarization, the gating field drives a strongly asymmetric modification of the flat bands: the flat valence band (FVB) evolves into an M-shaped dispersion at high field, whereas the flat conduction band (FCB) progressively flattens with increasing field. Comparison with calculations identifies critical parameters governing the band curvature of R5G, from which the resulting finite Berry curvature and near-ideal quantum geometry support the emergence of topological phases under electron doping at large fields. 
These results establish a direct link between the asymmetric phase diagram, band structure evolution, and quantum geometry, providing a microscopic framework for understanding correlated and topological phases in rhombohedral graphene.
\end{abstract}

\newpage

\renewcommand{\thefigure}{\textbf{Fig. \arabic{figure} $\bm{|}$}}
\renewcommand{\thetable}{\textbf{Table \arabic{table} $\bm{|}$}}
\setcounter{figure}{0}

Rhombohedral-stacked graphene (RG) exhibits a rich variety of correlated quantum phases under a displacement field, including correlated and Chern insulators\cite{WangFTrilayerNat19FM,WangFTrilayerNP19,Guorui_NatNano2024,Guorui_Science2024,Julong_Science2024,Julong_NatNano2024}, superconductivity\cite{WangFTrilayerNat19,YoungSC2021}, broken-symmetry metals and multiferroicity\cite{Young_Nature2021,Julong_Nature2023}, and most notably the fractional quantum anomalous Hall effect (FQAHE)\cite{Julong_Nature2024} and chiral superconductivity\cite{Julong_Nature2025Chiral}. Despite the rapid accumulation of experimental results, a unifying understanding of how these phases emerge upon doping and displacement field remains unclear. By compiling correlated phases reported in the literature\cite{Julong_Nature2024,Julong_Nature2025Chiral,WangFTrilayerNat19FM,WangFTrilayerNP19,Guorui_NatNano2024,Guorui_Science2024,Julong_Science2024,Julong_NatNano2024,WangFTrilayerNat19,Young_Nature2021,YoungSC2021,Julong_Nature2023,Xiaobo_NM2025,Julong_Nature2025,Young_Nature2025_SOC,Julong_NM2025_SC,Lu_PRL2026,Young_Nature2025,Liu_arXiv2025,Banerjee_arXiv2025,Liu_arXiv2025MF} into a unified phase diagram (Fig.~1), we identify a striking field-dependent electron-hole asymmetry that serves as an organizing principle for correlated phases in RG: while many correlated states are observed on the hole-doped side at small displacement fields ($D/\varepsilon_0$ $\lesssim$ 0.5 V/nm), the FQAHE\cite{Julong_Nature2024} has been reported only on the electron-doped side under large displacement fields ($D/\varepsilon_0$ $\approx$ 1 V/nm). 
Such a universal trend raises a fundamental question: why do the most exotic topological phases emerge exclusively on the electron-doped side under large displacement fields? More broadly, what microscopic mechanism governs the striking electron-hole asymmetry across the phase diagram?

\begin{figure*}[htbp]
	\centering
	\includegraphics[width=16 cm]{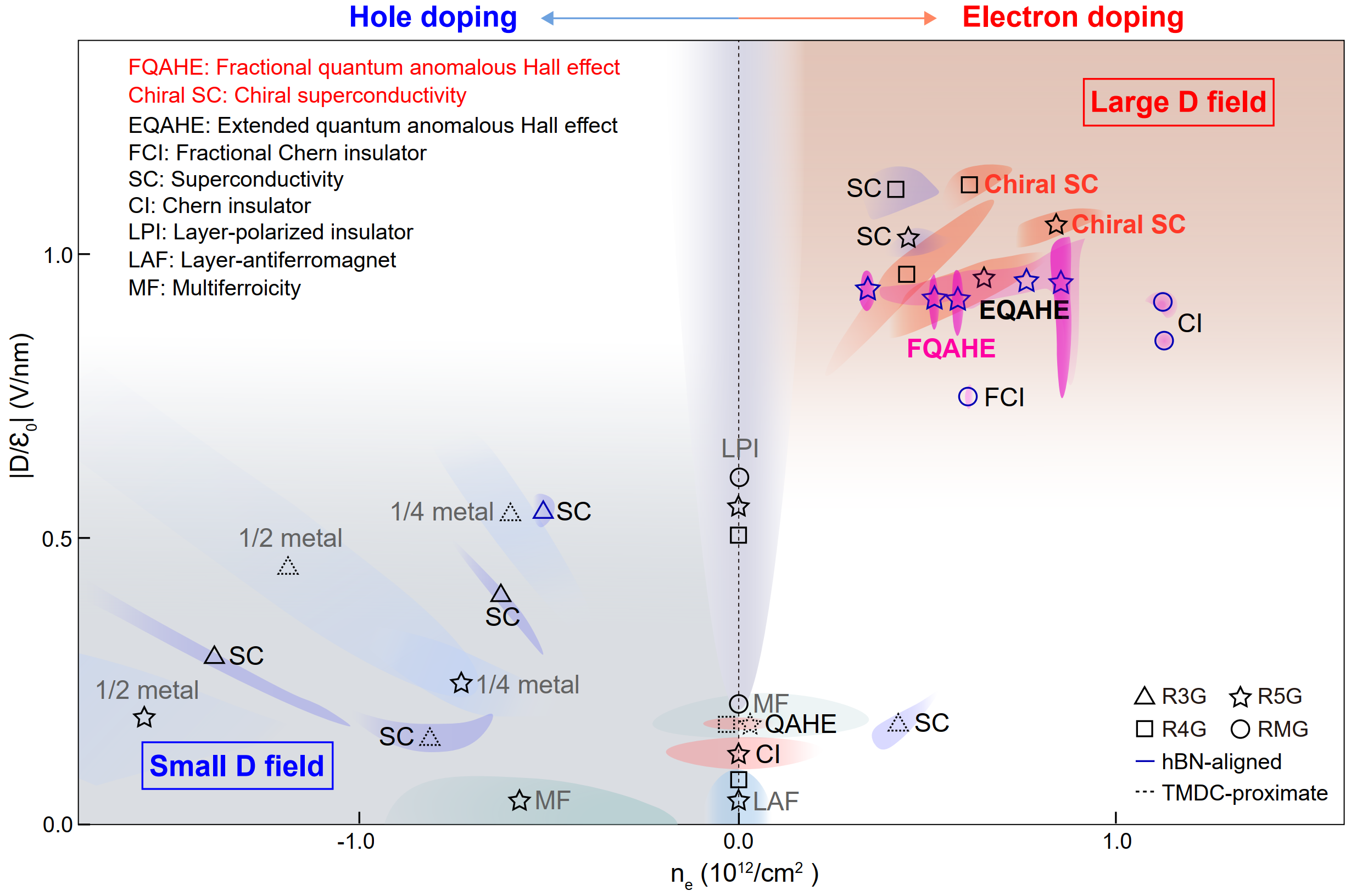}
	\caption{\textbf{A unified phase diagram of rhombohedral graphene by compiling correlated quantum phases reported in the literature.} Schematic summary of correlated quantum phases in rhombohedral graphene ranging from R3G (three layers) to RMG (six layers or thicker) upon application of displacement field ($D/\varepsilon_0$) and doping ($n_e$). Black markers represent few layer RG (non-aligned with BN), while blue and black dotted markers represent hBN-aligned and TMDC-proximitized samples, respectively.}
\end{figure*}

Rhombohedral graphene is distinguished by its unique interlayer stacking, which gives rise to two nearly degenerate flat bands at the Fermi level\cite{Koshino_PRB2009,MacDonaldPRB2010,Zhou_R5G_NM2025}. These flat bands are highly sensitive to an out-of-plane displacement field\cite{TaruchaNatNano2009,HeinzTriGNP2011,LauNatPhys2011}, making RG an exceptional platform for gate-tunable correlated and topological physics\cite{MacDonald_PRL2011,Falko_ComPhys2019,Guinea_NRP2023,Tan_PRX2024,Dong_PRB2024,JingWang_PRB2025,Xie_PRB2025,Bernevig_arXiv2025,shavit_PRB2026}. 
However, despite extensive theoretical and experimental efforts, the field-dependent electronic structure of RG has not been directly resolved. Transport measurements have revealed a rich landscape of correlated phases\cite{Julong_Nature2024,Julong_Nature2025Chiral,WangFTrilayerNat19FM,WangFTrilayerNP19,Guorui_NatNano2024,Guorui_Science2024,Julong_Science2024,Julong_NatNano2024,WangFTrilayerNat19,Young_Nature2021,YoungSC2021,Julong_Nature2023,Xiaobo_NM2025,Julong_Nature2025,Young_Nature2025_SOC,Julong_NM2025_SC,Lu_PRL2026,Young_Nature2025,Liu_arXiv2025,Banerjee_arXiv2025,Liu_arXiv2025MF}, providing important constraints on the underlying electronic structure, yet they do not directly resolve the momentum-resolved dispersion. Meanwhile, theoretical calculations depend sensitively on model details and the choice of parameters\cite{Koshino_PRB2009,MacDonaldPRB2010,MacDonald_PRL2011,Falko_ComPhys2019,Guinea_NRP2023,Tan_PRX2024,Senthil_PRL2024,Herzog-Arbeitman_2024_moire,JingWang_PRB2025,Dong_PRB2024,Xie_PRB2025,Bernevig_arXiv2025,shavit_PRB2026}. Within realistic parameter ranges, these models yield markedly different scenarios for the electronic structure evolution under displacement fields, ranging from nearly electron-hole symmetric dispersions\cite{Koshino_PRB2009,MacDonaldPRB2010} to strongly asymmetric ones\cite{Falko_ComPhys2019,Guinea_NRP2023,Tan_PRX2024,Senthil_PRL2024,Herzog-Arbeitman_2024_moire,JingWang_PRB2025}. This ambiguity underscores the need for direct spectroscopic determination of the field-dependent electronic structure.

The zero-field electronic structure of RG, including the effect of the moir\'e potential in aligned RG superlattice, was established in our recent work\cite{Zhou_R5G_NM2025}. That study, performed without a gate voltage, could not reveal the field-dependent electronic structure, which is fundamental for understanding the electron-hole asymmetric phase diagram under displacement fields.  Orthogonal to our previous work on zero-field electronic structure with a moir\'e potential, the present work introduces {\it in situ} electrostatic gating as an independent control axis, a degree of freedom fundamental to the entire phase diagram. This enables direct determination of the field-dependent electronic structure evolution, providing the basis for evaluating Berry curvature and quantum geometry in the high-field regime and offering microscopic insight into the unified phase diagram.

Here we directly visualize the displacement-field induced band evolution in rhombohedral pentalayer graphene (R5G), by using nanospot angle-resolved photoemission spectroscopy (NanoARPES) with \textit{in situ} electrostatic gating. 
Our measurements reveal a pronounced asymmetric response between the flat conduction band (FCB) and flat valence band (FVB). As the gate voltage increases, the FVB develops a distinct M-shaped dispersion, while the FCB progressively flattens, accompanied by gap opening and spectral-weight redistribution indicative of field-induced layer polarization. Combining experimental results with theoretical calculations, we show that large fields generate finite Berry curvature and near-ideal quantum geometry. Notably, the flattest FCB and most ideal quantum geometry are both achieved near an optimal field, consistent with the regime where the FQAHE has been observed\cite{Julong_Nature2024}. Our results establish a direct spectroscopic link between the transport phase diagram, electronic structure and quantum geometry, providing a unified microscopic framework for understanding correlated phenomena in RG systems.

{\bf An overview of gating effects on the electronic structures of R5G}

\begin{figure*}[htbp]
	\centering
	\includegraphics[width=16 cm]{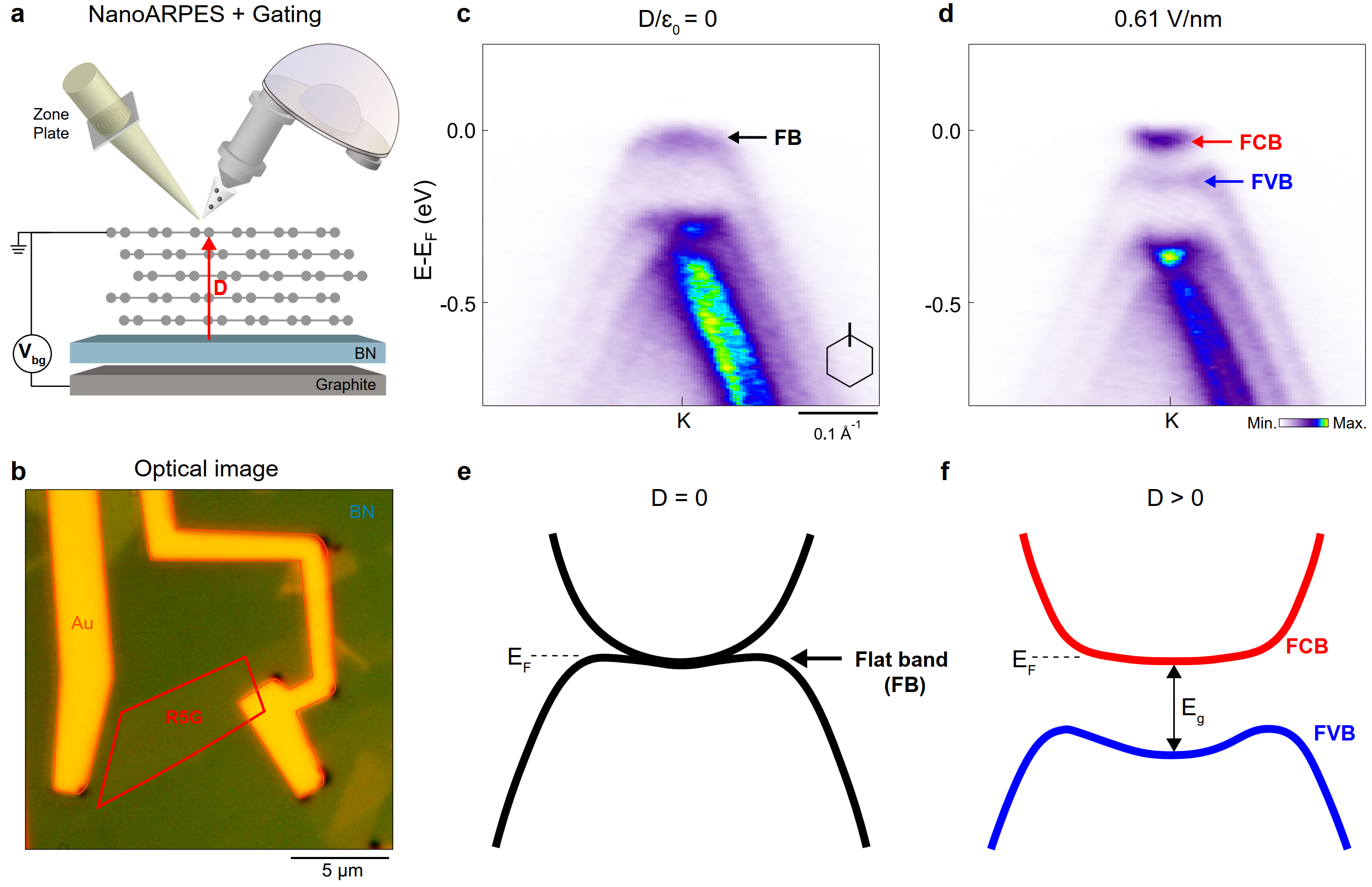}
	\caption{\textbf{Experimental electronic structures of R5G under a large displacement field.} 
	\textbf{a}, Schematic of NanoARPES measurements with \textit{in situ} gating $V_{bg}$, which applies a displacement field to the sample and dopes it. \textbf{b}, Optical image of the R5G device (sample S1), where R5G, BN and Au electrodes are marked by red, blue and orange, respectively. \textbf{c,d}, Dispersion images  measured along the $\Gamma$-K direction at $D/\varepsilon_0$ = 0 (panel \textbf{c}) and $D/\varepsilon_0$ = 0.61 V/nm (panel \textbf{d}). \textbf{e,f}, Schematic summary of electronic structures in (\textbf{c}) and (\textbf{d}).}
\end{figure*}

To explore the effect of displacement field on the flat bands of R5G, we fabricated high-quality R5G devices, optimized for NanoARPES measurements with \textit{in situ} electrostatic gating (schematic in Fig.~2a). Since NanoARPES measurements require exposure of the top surface\cite{Wilson_Nature2019,Gate_NanoLett2019,ZhangHY_NC2024}, a gate voltage  is applied only to the backside ($V_{bg}$), which induces both a effective displacement field $D/\varepsilon_0$ = $\varepsilon_rV_{bg}/2d$ and carrier doping into the sample (see Methods), where $\varepsilon_r$ = 4 and $d$ = 26.4 nm are the dielectric constant and thickness of BN. Figure 2b shows an optical image of one such R5G device (sample S1, see Methods and Extended Data Fig.~1,2 for more details of sample S1 and S2). In our experiments, a large displacement field up to $D/\varepsilon_0$ = 1.18 V/nm can be applied {\it in situ} (see Methods), enabling direct access to the strong-field regime where FQAHE and chiral superconductivity are observed (see phase diagram in Fig.~1).

Figure~2c,d compares the electronic structure of R5G measured along the $\Gamma$-K direction at $D/\varepsilon_0$ = 0  ($V_{bg}$ = 0) and $D/\varepsilon_0$ = 0.61 V/nm ($V_{bg}$ = 8 V). Before gating, a flat band is observed near the Fermi energy (black arrow in Fig.~2c), where the FVB and FCB are degenerate near the K point, as schematically shown in Fig.~2e. Applying gating voltages induces three major changes. First, the displacement field lifts the degeneracy of FCB and FVB, opening a gap $E_g$ at the K point (see Extended Data Fig.~3 for a detailed analysis), as schematically illustrated in Fig.~2f. Second, the FCB shows stronger intensity than the FVB, serving as a clear spectral signature of the layer polarization, as discussed in detail below. Third, the band curvatures of both flat bands near the K point change significantly, with the FVB developing an M-shaped dispersion while the FCB becomes flatter. In the following, we focus on the evolution of the gap opening, layer-polarization, band curvature, and the topological properties of FCB and FVB upon gating, which together unveil the rich physics of R5G under strong displacement fields.

{\bf Gap opening and spectroscopic signatures of layer polarization upon gating}

\begin{figure*}[htbp]
	\centering
	\includegraphics[width=16 cm]{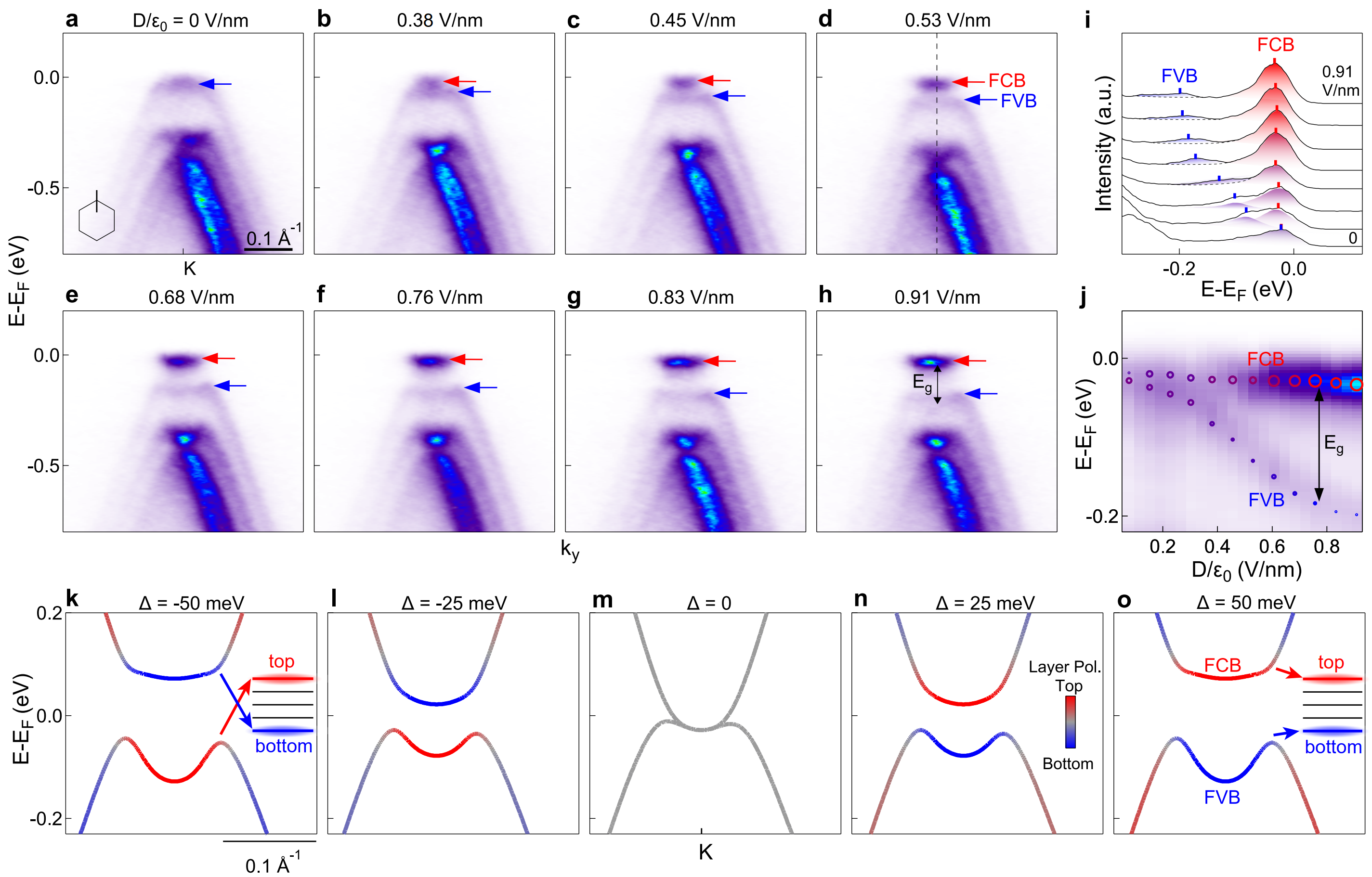}
	\caption{\textbf{Evolution of gap size and spectroscopic evidence of layer polarization upon gating.} 
	\textbf{a-h}, Dispersion images measured on sample S1 along the $\Gamma$-K direction as $D/\varepsilon_0$ increases gradually from 0 to 0.91 V/nm. Red and blue arrows mark the FCB and FVB, respectively. \textbf{i}, EDCs at the K point (marked by dashed line in \textbf{d}) at different displacement fields. The spectral weight transfer between the FCB (red shading) and FVB (blue shading) indicates the layer polarization. \textbf{j}, Intensity map showing EDC at the K point at different displacement fields. The size of red and blue markers represents the spectral weights of the FCB and FVB, respectively, and their energy separation defines the gap size $E_g$. \textbf{k-o}, Calculated dispersions and layer polarizations along the $\Gamma$-K direction with interlayer potential difference $\Delta$ = -50, -25, 0, 25, 50 meV respectively. Red-gray-blue colors indicate the projected spectral weight to the top and bottom layers.}
\end{figure*}

Figure~3a-h shows the evolution of electronic structures of R5G as the displacement field $D/\varepsilon_0$ increases gradually from 0 to 0.91 V/nm. At $D/\varepsilon_0$ = 0, the FVB is observed near the Fermi energy (blue arrow in Fig.~3a). At $D/\varepsilon_0$ = 0.38 V/nm, the flat band clearly splits into the FCB and FVB with comparable spectral intensity (red and blue arrows in Fig.~3b). The FCB becomes partially occupied and appears as a faint dot-like feature, while the FVB shifts down in energy due to the electron doping. Further increasing $D/\varepsilon_0$ gradually separates the FCB and FVB at the K point, indicating a larger gap opening. Analysis from energy distribution curves (EDCs) and the corresponding image plot at the K point (Fig.~3i,j) shows that the gap size $E_g$ increases with displacement fields, reaching $E_g$ = 0.17 $\pm$ 0.02 eV at $D/\varepsilon_0$ = 0.91 V/nm. In addition to the gap opening, gating also induces a pronounced spectral weight transfer from the FVB to FCB. As $D/\varepsilon_0$ increases, the FCB dominates the low-energy spectrum (red arrows in Fig.~3c-h), while the FVB shifts to higher binding energy and diminishes in intensity, as revealed in the EDCs and image in Fig.~3i,j. Such electron-hole-asymmetric intensity evolution is a spectroscopic signature of the field-induced layer polarization. 

The field-induced gap opening and layer polarization are supported by theoretical calculations in Fig.~3k-o. In ungated R5G, the flat bands are gapless and layer-symmetric, with FCB and FVB touching at the K point. Introducing an interlayer potential difference ($\Delta$ = 25 and 50 meV, Fig.~3n,o) reproduces key experimental features, including the gap opening and the layer polarization. In particular, the FCB localizes predominantly on the top graphene layer (indicated by red color in Fig.~3o), while the FVB localizes on the bottom layer (blue) upon positive gating. Reversing the gate polarity switches the layer polarization (Fig.~3k), accompanied by hole doping, consistent with NanoARPES measurements at negative gating (Extended Data Fig.~4). 

The localization of the FCB to the top layer under positive gating provides a natural explanation for the enhanced spectral weight observed in NanoARPES, considering its surface sensitivity. Such field-induced layer-polarization introduces a tunable degree of freedom, forming a key microscopic basis for electrically controlled correlated phases in RG, including the QAHE with TMDC-proximity\cite{Guorui_Science2024,Julong_Science2024}, FQAHE and Chern insulating states in the moir\'e-distant side\cite{Julong_Nature2024,Xiaobo_NM2025,Julong_Nature2025,Liuxiaoxue_arXiv2026}.

{\bf Field-tunable band curvatures and asymmetric flattening of FCB and FVB}

\begin{figure*}[htbp]
	\centering
	\includegraphics[width=16 cm]{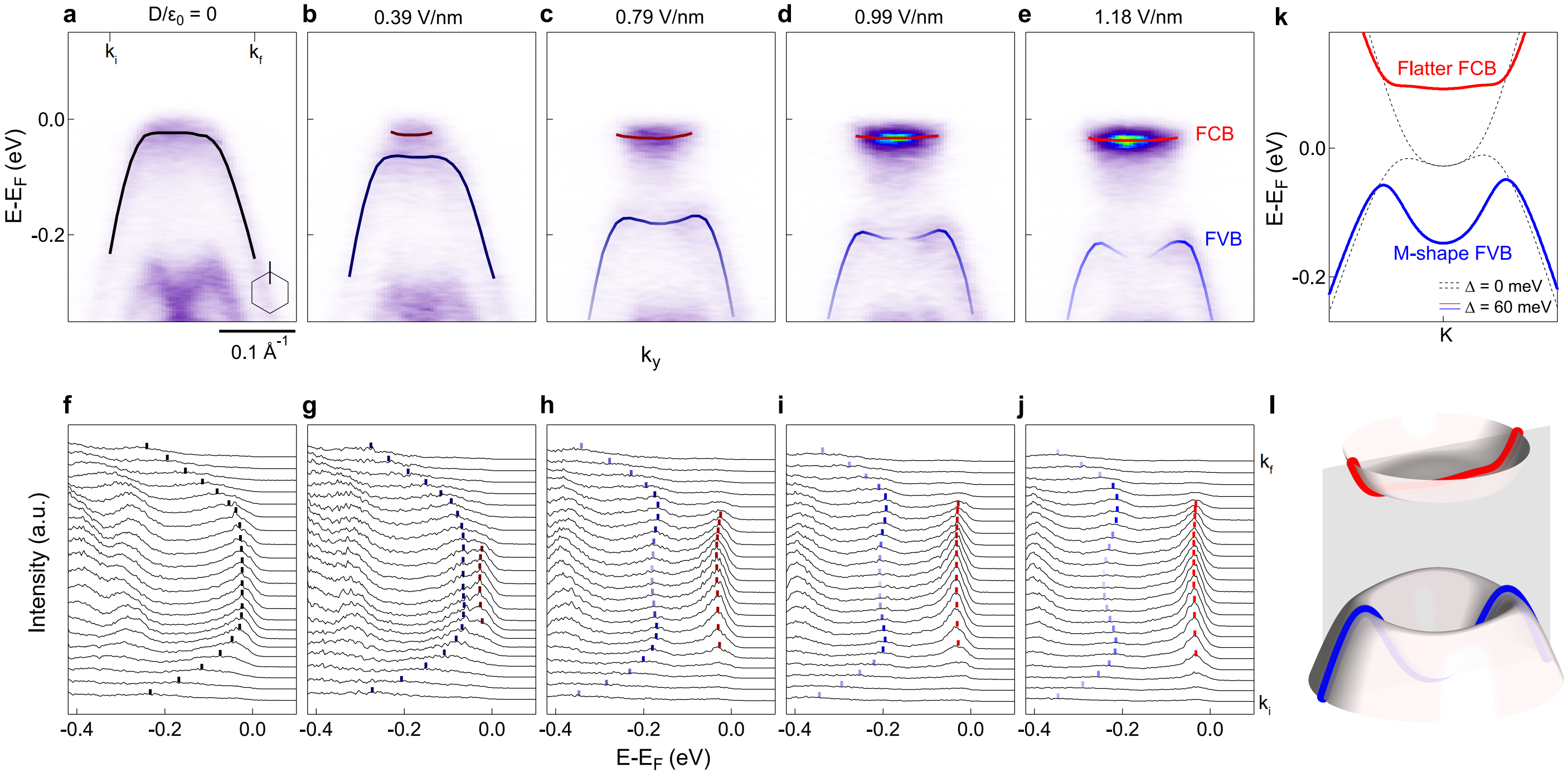}
	\caption{\textbf{Field-tunable band curvatures and asymmetric evolution of FCB and FVB.} \textbf{a-e}, Dispersion images measured along the $\Gamma$-K direction at $D/\varepsilon_0$ = 0, 0.39, 0.79, 0.99, and 1.18 V/nm on sample S2. Colored curves trace the dispersions of the FCB and FVB. \textbf{f-j}. EDCs extracted from \textbf{a-e} at momenta $k_i$ to $k_f$ (labeled in \textbf{a}), where colored ticks denote the peak positions. \textbf{k}, Calculated dispersions along the $\Gamma$-K direction for $\Delta$ = 0 (gray dotted curves) and $\Delta$ = 60 meV (blue and red curves). \textbf{l}, Calculated three-dimensional electronic structure for $\Delta$ =  60 meV.}
	\label{Fig4}
\end{figure*}

The band curvature, in particular its flatness, plays an important role in electronic correlations and topological properties, which can be finely manipulated by the displacement fields. Figure 4a-e shows zoomed-in NanoARPES spectra at a few representative displacement fields, revealing a pronounced asymmetric modulation of the FCB and FVB band curvatures. With increasing gate voltage, the FCB flattens while maintaining a strong spectral weight near the Fermi level. In contrast, the FVB evolves into a distinct M-shaped dispersion, featuring a dip at the K point and shoulders on both sides. By following peak positions in EDCs in Fig.~4f-j, the extracted dispersions are overplotted as colored curves in Fig.~4a-e, highlighting the strong electron-hole asymmetric electronic structure in the strong-field regime. 

Specifically, the FCB develops an increasingly flat band bottom as the band bottom spans larger momentum range upon gating (dark red to red curves in Fig.~4a-e), enhancing its density of states and electronic correlation, whereas the FVB becomes clearly M-shaped (dark blue to blue curves in Fig.~4a-e). Such marked electron-hole asymmetry suggests that strong correlation effects are more easily realized under the electron doping region when the Fermi level intersects the flattened FCB. This asymmetric band curvature modulation underlies the rich field-driven phase diagram, offering an intuitive picture of why the FQAHE and chiral superconductivity favor electron-doped R5G at large displacement fields.

Figure 4k shows calculated dispersions at $\Delta$ = 0 and 60 meV, using experimentally extracted hopping parameters\cite{Zhou_R5G_NM2025}, showing good agreement with the dispersions measured at $D/\varepsilon_0$ = 0 and 1.18 V/nm (see Extended Data Fig.~5). The calculations reproduce nicely the strong asymmetric reshaping of the FCB and FVB curvatures observed experimentally. Such reshaping of the flat bands is dominantly determined by the interlayer skewed hopping $\gamma_4$ and the intrinsic inversion-symmetric potential $V_{ISP}$ (Extended Data Fig.~6). The flattened FCB in the strong field regime naturally supports the emergence of FQAHE ($D/\varepsilon_0$ $\approx$ 1.0 V/nm) in the electron-doped side, where enhanced correlations in flat bands are critical. Meanwhile, the field-induced modification of the FVB generates an annular Fermi surface (Extended Data Fig.~7), which has been proposed to promote unconventional superconductivity via the Kohn-Luttinger mechanism\cite{Kohn_PRL1965,PRB2010,Berg_PRL2021,YoungSC2021}.

\begin{figure*}[htbp]
	\centering
	\includegraphics[width=16 cm]{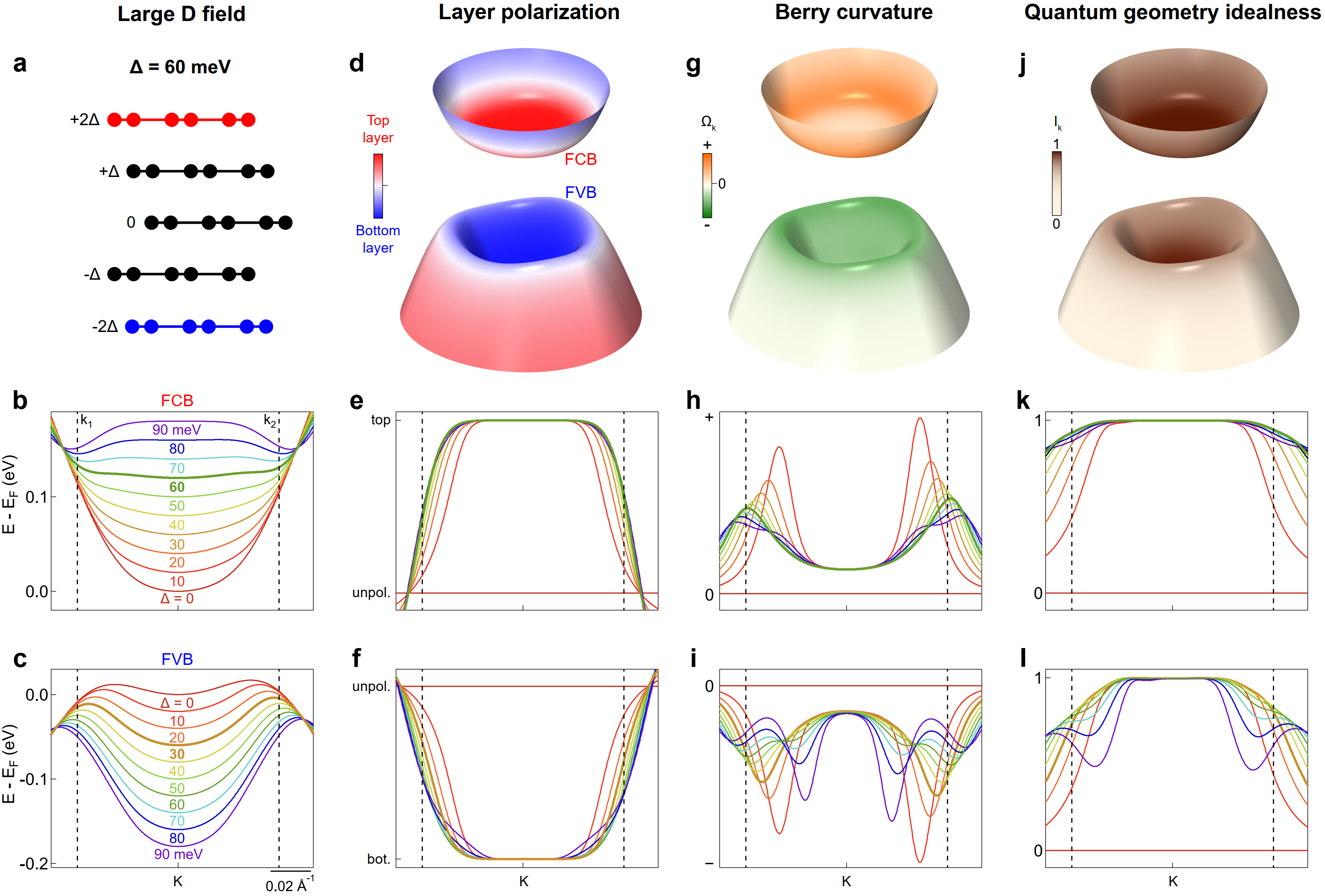}
	\caption{\textbf{Field-induced layer polarization, Berry curvature and quantum geometry idealness.} 
	\textbf{a}, Schematics of R5G with interlayer potential difference $\Delta$ = 60 meV. \textbf{b,c}, Calculated dispersion of FCB and FVB along the $\Gamma$-K direction with $\Delta$ from 0 to 90 meV.
	\textbf{d}, Calculated three-dimensional electronic structure with $\Delta$ = 60 meV, where red and blue colors correspond to the top and bottom layer polarizations respectively. \textbf{e,f}, Calculated layer polarization of FCB and FVB with $\Delta$ from 0 to 90 meV. \textbf{g}, Calculated three-dimensional Berry curvature map at $\Delta$ = 60 meV. \textbf{h,i}, Calculated Berry curvature of FCB and FVB with $\Delta$ from 0 to 90 meV. \textbf{j},  Calculated three-dimensional quantum geometry idealness map at $\Delta$ = 60 meV. \textbf{k,l}, Calculated quantum geometry idealness of FCB and FVB with $\Delta$ from 0 to 90 meV. Momenta $k_1$ to $k_2$ indicate the flat band region with a momentum separation of 0.1 $\text {Å}$$^{-1}$.}
	\label{Fig5}
\end{figure*}

{\bf Field-induced Berry curvature and ideal quantum geometry}

NanoARPES measurements on gated R5G devices show that a displacement field profoundly reshapes the flat-band electronic structure, driving a transition from layer-degenerate, gapless flat bands to gapped, layer-polarized FCB and FVB, as supported by calculated results in Fig.~5. These key experimental features, including the progressive flattening of the FCB and the M-shaped FVB, are well reproduced by theoretical calculations incorporating an interlayer potential difference $\Delta$. The FCB and FVB are evaluated within a momentum window (indicated by vertical lines in Fig.~5b,c), corresponding to the topological surface states in the bulk limit\cite{Zhou_PNAS2024}. This range is set by the hopping parameters by $\Delta k \approx 4\gamma_0/(\sqrt{3}at_0)$ = 0.1 $\text {Å}^{-1}$, where $t_0$ and $\gamma_0$ are the in-plane and out-of-plane nearest neighbour hopping parameters and $a$ is the lattice constant. Within this momentum window, the FCB and FVB also become predominantly localized on the top and bottom layers respectively with increasing displacement field (Fig.~5e,f).

Such displacement-field control of the band dispersion and layer character fundamentally reshapes the band topology. This effect can be further quantified by the Berry curvature $\Omega_k$, given by the imaginary part of the quantum geometric tensor\cite{NiuQian_RMP2010,AkshayPRB2013,Bernevig_NRP2022,
Xie_NSR2024,GregoryPRB2024,Cano_PRX2025,Torma_PRB2026,Mathias_arXiv2025}, $\mfk{g}_{ij}(\kk)$. In the absence of a displacement field, the flat bands exhibit vanishing Berry curvature. Upon introducing $\Delta$ = 60 meV (see Extended Data Fig.~8 for additional $\Delta$ values), the system enters a regime in which the FCB and FVB acquire Berry curvature of opposite signs near the gap edge (Fig.~5g). At such large displacement fields, the Berry curvature of the FCB becomes nearly uniform in most of the flat band momentum region. 
The uniform Berry curvature leads to ideal quantum geometry\cite{Rahul_2014_band,Torma_PRB2026}, which is defined by $I_\kk = \frac{|\Omega_{\kk}|}{\mrm{Tr} g(\kk)}$ =1, where $g(\kk)$ is the quantum metric (see Methods).

By systematically comparing calculated results across different $\Delta$ values, we further pinpoint critical field conditions governing the band curvature and topological properties of R5G. At $\Delta$ = 60 meV (corresponding to the experimental result at displacement field $D/\varepsilon_0$ = 1.18 V/nm), the FCB becomes maximally flat (Fig.~5b) within its drumhead region while exhibiting nearly ideal quantum geometry (Fig.~5k), forming the ideal condition to host both correlated and topological phases. Fields that deviate from such a condition, whether too weak or too strong, result in a significant increase of the FCB bandwidth and a degradation of the quantum geometry idealness. Strikingly, this optimal condition matches closely with the experimental displacement fields regime of FQAHE and chiral superconductivity ($D/\varepsilon_0$ $\sim$ 1 V/nm). In contrast, although the FVB is relatively flat near zero displacement field (Fig.~5c), achieving ideal quantum geometry requires a finite displacement field, for example, $\Delta$ = 30-50 meV (Fig.~5l). This mismatch precludes the simultaneous optimization of both electron correlations and nontrivial band topology on the hole-doped side. Such intrinsic asymmetry provides an intuitive microscopic explanation for the emergence of FQAHE in the electron-doped FCB under large displacement fields. By directly resolving the band dispersions and combining them with theory, we establish a microscopic electronic structure basis for these exotic phases.

{\bf Summary and Discussion}

Our results provide direct visualization of displacement-field-driven asymmetric band modification, revealing the microscopic origin of the pronounced electron-hole asymmetric phases observed in RG systems, which distinguishes RG from other flat band platforms such as twisted bilayer graphene\cite{Gordon_Science2019,Yankowitz_Science2019,Yazdani_NRM2024}. In particular, the emergence of a flattened FCB at high displacement fields explains why exotic correlated phases in the electron-doped side require large displacement fields, whereas the flat FVB is preserved only at small fields, consistent with the stabilization of correlated phases on the hole-doped side at much lower displacement fields. Our work shows that displacement field fundamentally reshapes the flat-band dispersion, modifies layer character and topology in RG, thereby providing a direct microscopic explanation for its asymmetric phase diagram and the emergence of FQAHE and chiral superconductivity. These findings resolve a fundamental question in the field and establish a new paradigm for designing fractional topological states through band structure control. 

\newpage

\begin{methods}

\renewcommand{\thefigure}{\textbf{Extended Data Fig.~\arabic{figure} $\bm{|}$}}
\setcounter{figure}{0}

\subsection{Sample preparation and characterization.}

The exfoliation, identification, and isolation of R5G followed procedures described in our previous work\cite{Zhou_R5G_NM2025}. Gate-compatible devices were fabricated using a dry transfer technique. The graphite back gate, middle BN, and R5G were sequentially picked up with a polypropylene carbonate (PPC) film and transferred onto SiO$_2$/Si substrate, leaving the R5G fully exposed on top. Electrical contacts to the ground and back-gate electrodes were patterned by standard e-beam lithography followed by Cr/Au deposition. The relative angle between R5G and BN was intentionally kept away from 0° to avoid moir\'e effects. The BN thicknesses were 26.4 nm (sample S1) and 20.3 nm (sample S2). The displacement field was defined as $D/\varepsilon_0$ = $\varepsilon_rV_{bg}/2d$, where $d$ is the BN thickness and $\varepsilon_r$ = 4 is its dielectric constant. In our back-gated geometry, the displacement field is not independently tunable but is intrinsically coupled to the carrier density. To have a direct comparison with dual-gated transport results, we adopt a conventionally defined effective displacement field following the dual-gate formalism\cite{Guorui_NatNano2024}. Within this framework, the graphene layer is treated as an equipotential reference, and here D provides a consistent measure to track the field-dependent evolution of the flat bands. While this approximation does not fully capture screening effects and device asymmetry, the overall trends upon gating remain robust. Data in Fig.~2,3 and Extended Data Fig.~1,3,4 were obtained from sample S1, while Fig.~4 and Extended Data Fig.~2,5,6 were obtained from sample S2, which gave the same results in the overlapping field regions.

\subsection{ARPES measurements.}

NanoARPES measurements were performed at the ANTARES beamline of Synchrotron SOLEIL in France at the temperature of 78 K, using a photon energy of 95 eV and an overall energy resolution of 40 meV. Data in Extended Data Fig.~4 and Extended Data Fig.~7a were collected at beamline I05 of Diamond Light Source in the UK at the temperature of 20 K, using a photon energy of 75 eV and an energy resolution better than 33 meV. The beam spot size was smaller than $\SI{1}{\micro\meter}$. Before NanoARPES measurements, R5G samples were annealed at 180$~^\circ \text{C}$ for several hours in ultrahigh vacuum until clear NanoARPES images and sharp dispersions were obtained (Extended Data Fig.~1,2). During measurements, the R5G device was electrically grounded. The displacement field was applied via a bottom-gate voltage using a Keithley source meter, with the static leakage current maintained below 1 nA. The beam-induced voltage drop was estimated to be below 0.25 V (corresponding to \textless~0.02 V/nm) throughout the NanoARPES measurement.

\subsection{Theoretical calculations.}

For pristine R5G, we adopt the tight-binding model to capture its electronic structure at one valley, 
\begin{align} 
	H_{0} &= \sum_{\mathbf{k}} \sum_{l=0}^{4} \sum_{\alpha,\alpha'} c^\dagger_{\mathbf{k},l,\alpha} \begin{pmatrix}
		V_l & \gamma_0 f(\mathbf{k}) \\
		\gamma_0f^*(\mathbf{k}) & V_l \\
	\end{pmatrix}_{\alpha,\alpha'}  c_{\mathbf{k},l,\alpha'} \\\nonumber
	&+ \sum_{\mathbf{k}} \sum_{l=0}^{3} \sum_{\alpha, \alpha'} c^\dagger_{\mathbf{k},l,\alpha} \begin{pmatrix}
		- \gamma_4 f(\mathbf{k})& - \gamma_3 f^*(\mathbf{k}) \\
		\gamma_1 & - \gamma_4 f(\mathbf{k}) \\
	\end{pmatrix}_{\alpha,\alpha'} c_{\mathbf{k},l+1,\alpha'} + \mathrm{H.c.} \\\nonumber
	&+ \sum_{\mathbf{k}} \sum_{l=0}^{2} \sum_{\alpha,\alpha'} c^\dagger_{\mathbf{k},l,\alpha} \begin{pmatrix}
		0 & \gamma_2 \\
		0 & 0 \\
	\end{pmatrix}_{\alpha,\alpha'} c_{\mathbf{k},l+2,\alpha'} + \mathrm{H.c.}
\end{align}
The index $\alpha=A,B$ denotes graphene sublattices, and $l =0, 1, \dots, 4$ denote layers from bottom to top, and the momentum $\mathbf{k}$ is measured from the $\mathbf{K}$ point of R5G Brillouin zone.
The $c^\dagger_{\mathbf{k},l,\alpha}$ ($c_{\mathbf{k},l,\alpha}$) are electron creation (annihilation) operators. The tight-binding model and hopping parameters $\gamma_0, \gamma_1, \dots, \gamma_4$ (schematically shown in Extended Data Fig.~5a) are adopted to follow our previous results\cite{Zhou_R5G_NM2025}, in order to recover the band structures of pristine R5G. The factor $f(\mathbf{k}) \approx -\frac{\sqrt{3}}{2} a_0 k_- + \frac{1}{8} (a_0 k_+)^2$ expands to the second order of $k$, where $k_\pm=k_x\pm ik_y$. The potential on the $l$ layer, $V_l = V_{\rm ISP} |2-l| + \Delta*(l-2)$, where $V_{\rm ISP}$ describes the intrinsic inversion-symmetric potential, and $\Delta$ is the gate-induced potential difference between adjacent layers. The wavefunctions at momentum $\mathbf{k}$ is denoted as $\left| \mathbf{k} \right\rangle  = \sum\limits_{l,\alpha } {{\lambda _{l\alpha }}\left| {l\alpha } \right\rangle }$. Layer polarization is defined as $P = \frac{{{{\left| {{\lambda _{4A}}} \right|}^2} + {{\left| {{\lambda _{4B}}} \right|}^2}}}{{\left( {{{\left| {{\lambda _{0A}}} \right|}^2} + {{\left| {{\lambda _{0B}}} \right|}^2}} \right) + \left( {{{\left| {{\lambda _{4A}}} \right|}^2} + {{\left| {{\lambda _{4B}}} \right|}^2}} \right)}}$. Extended Data Fig.~5b,c shows the comparison of calculated results at $\Delta$ = 0 and 60 meV with experimental results at $D/\varepsilon_0$ = 0 and 1.18 V/nm, which shows overall good agreement. Here the asymmetric band curvature of FCB and FVB mainly comes from the interlayer skewed hopping $\gamma_4$ and the intrinsic inversion-symmetric potential $V_{ISP}$, as shown in Extended Data Fig.~6.

For a Bloch band $|\kk\rangle$, the complex quantum geometric tensor is defined by
\begin{align}  \label{eq:gij}
	\mfk{g}_{ij}(\kk) &= \Big(\partial_{k_i} \langle \kk | \Big)  \Big( 1 - | \kk \rangle\langle \kk |  \Big) \Big( \partial_{k_j} | \kk \rangle  \Big)
\end{align}
where $i,j = x,y$. 
Especially, the Berry curvature is given by $\Omega_{\kk} = -2\mrm{Im} \mfk{g}_{xy}(\kk) = 2\mrm{Im} \mfk{g}_{yx}(\kk) = \ii \Big(\partial_{k_x}\langle \kk| \Big) \Big(\partial_{k_y} | \kk \rangle \Big) - \ii \Big(\partial_{k_y}\langle \kk| \Big) \Big(\partial_{k_x} | \kk \rangle \Big)$, and the Fubini-Study quantum metric is given by $\mrm{Re} \mfk{g}_{ij}(\kk) = g_{ij}(\kk)$. 
Equation (2) by construction must obey a trace inequality, $\mrm{Tr} g(\kk) \ge |\Omega_{\kk}|$\cite{Rahul_2014_band}. For $\Delta=0$, R5G obeys inversion symmetry $P$ besides the time-reversal symmetry $T$. 
For a non-degenerate Bloch state $|\kk\rangle$, $PT$ will enforce the top and bottom layer components to be equal, and its Berry curvature to vanish. 

When the trace inequality is saturated, and the distribution of Berry curvature tends to be uniform, the band is said to be ideal for realizing fractional Chern insulators\cite{WenXGPRL2011,SarmaPRL2011}, as the projected interaction operators will closely follow those in the lowest Landau level\cite{Rahul_2014_band, Liu_2024_recent}. We quantify the wellness of the trace condition by the idealness $I_\kk = \frac{\Omega_{\kk}}{\mrm{Tr} g(\kk)}$. The properties of FVB and FCB in the range of $|k| \ll \frac{\gamma_1}{v_F}$ can be understood in the infinite-layer limit, where $v_F$ is Dirac velocity, and $\gamma_1$ is the perpendicular hopping in adjacent layers. There, these flat bands originate as the drumhead surface states of a bulk RG that forms the Dirac nodal line semimetal\cite{Zhou_PNAS2024}, and adopt simple ansatz $|k, top\rangle = \sum_{l} \left(  \frac{v_F (k_x-ik_y)}{\gamma_1}  \right)^{n-l} |k,l,B\rangle$ and $|k, bottom\rangle = \sum_{l} \left(  \frac{v_F (k_x+ik_y)}{\gamma_1}  \right)^{l} |k,l,A\rangle$ up to normalization, where $l$ labels the layers. The two surface states decrease exponentially into the bulk, hence are kinematically decoupled, while $D$ field splits their energy. They are (anti-)holomorphic function of $k_x-ik_y$ or $k_x+ik_y$, respectively, hence saturate the idealness, with Berry curvature $\Omega_{\kk}$ = $\frac{1}{\left( 1 - \left| \frac{v_F k}{\gamma_1} \right|^2 \right)^2}$. As $|k|$ increases, the surface states extend further into the bulk, so that the finite-layer effect starts to appear.

Figure 5 shows that at a high displacement field of $\Delta$ = 60 meV, the FCB is extremely flat at the drumhead region (labeled as $k_1$ and $k_2$), while the quantum geometry becomes ideal, favoring exotic correlated phenomena such as the FQAHE and chiral superconductivity.

\clearpage
\appendix

\newpage
	
\renewcommand{\thefigure}{\textbf{Extended Data Fig.~\arabic{figure} $\bm{|}$}}
\setcounter{figure}{0}

\begin{figure*}[htbp]
	\centering
	\includegraphics[width=15cm]{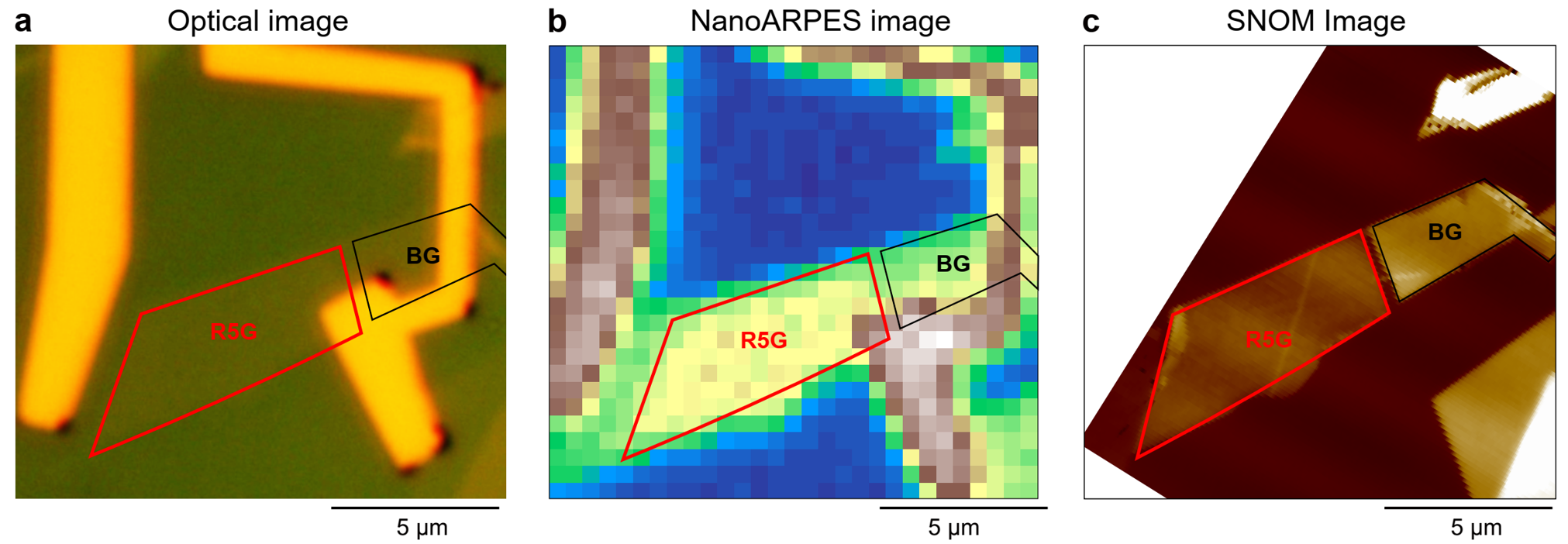}
	\caption{\textbf{Characterization of R5G samples.} \textbf{a}, Optical image of the R5G sample S1. \textbf{b}, NanoARPES spatial image to locate the R5G region. \textbf{c}, Infrared optical image to confirm the rhombohedral stacking of R5G (red curve).}
	\label{SI_Fig1}
\end{figure*}

\begin{figure*}[htbp]
	\centering
	\includegraphics[width=15cm]{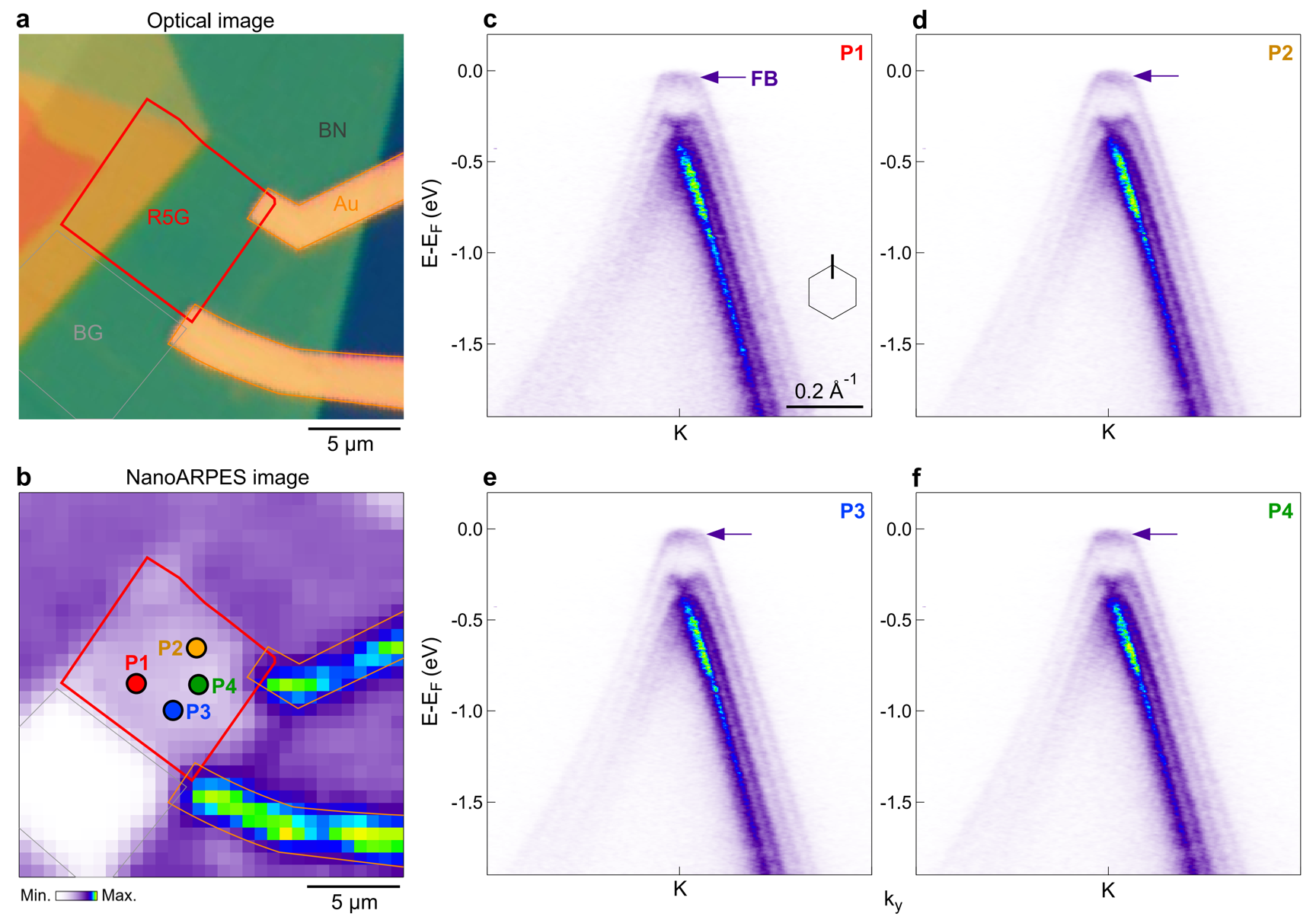}
	\caption{\textbf{Homogeneity of electronic structure among the R5G sample.}
	\textbf{a}, Optical image of R5G sample S2, where
	R5G, Bernal stacked graphene and Au electrodes are marked by red, gray, and orange curves. \textbf{b}, Spatially-resolved NanoARPES intensity map at the same region of the optical image. \textbf{c-f}, Dispersion images along $\Gamma$-K measured at positions from P1 to P4 as indicated in \textbf{b}.}
	\label{SI_Fig2}
\end{figure*}

\begin{figure*}[htbp]
	\centering
	\includegraphics[width=16cm]{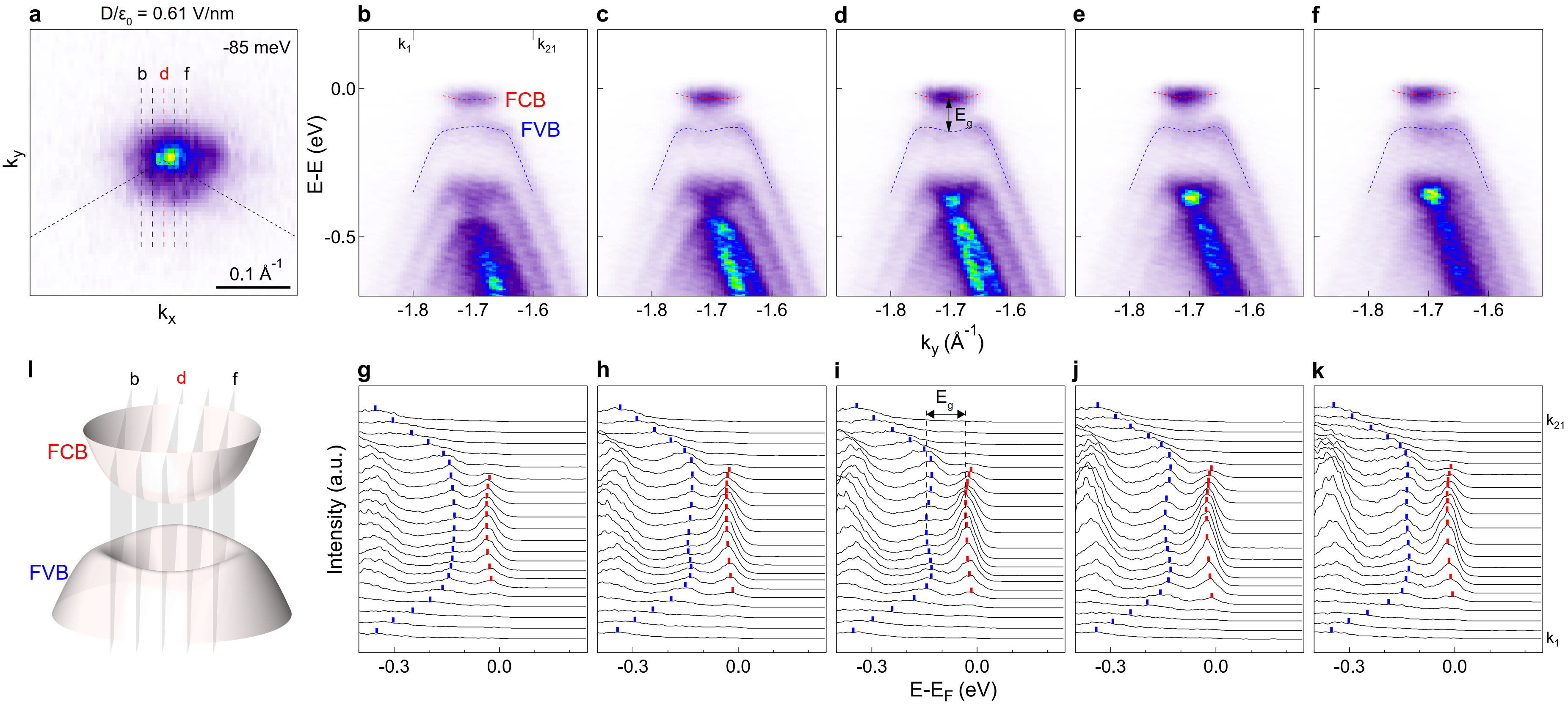}
	\caption{\textbf{Gap opening between FCB and FVB at effective displacement field of 0.61 V/nm (sample S1).} \textbf{a}, Energy contour measured at -85 meV. \textbf{b-f}, Parallel cuts measured along dashed lines in \textbf{a}. Panel \textbf{d} cuts through the K point as indicated by the red dashed line in \textbf{a}. \textbf{g-k}, EDCs extracted from momenta $k_1$ to $k_{21}$ from \textbf{b-f}. \textbf{l}, Schematic to show the gap opening between FCB and FVB.}
	\label{SI_Fig3}
\end{figure*}

\begin{figure*}[htbp]
	\centering
	\includegraphics[width=16cm]{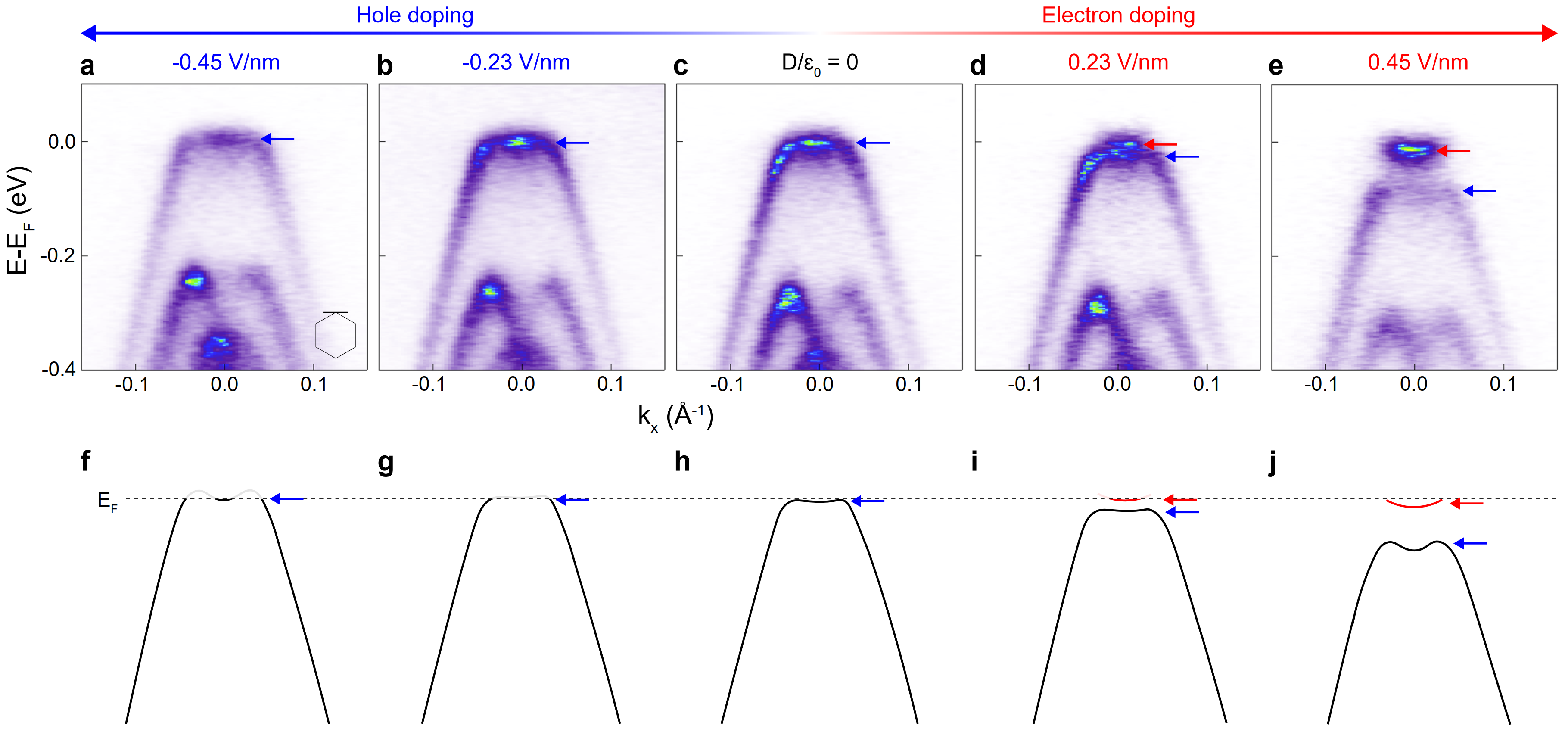}		\caption{\textbf{Band evolution upon negative and positive gating.} \textbf{a-e}, NanoARPES dispersion images with effective displacement field changes from -0.45 V/nm to 0.45 V/nm (sample S1). The cutting direction is perpendicular to $\Gamma$-K as indicated by the inset. \textbf{f-j}, Schematic drawings to show the fillings with holes and electrons under negative and positive gating.}
	\label{SI_Fig4}
\end{figure*}

\begin{figure*}[htbp]
	\centering
	\includegraphics[width=15cm]{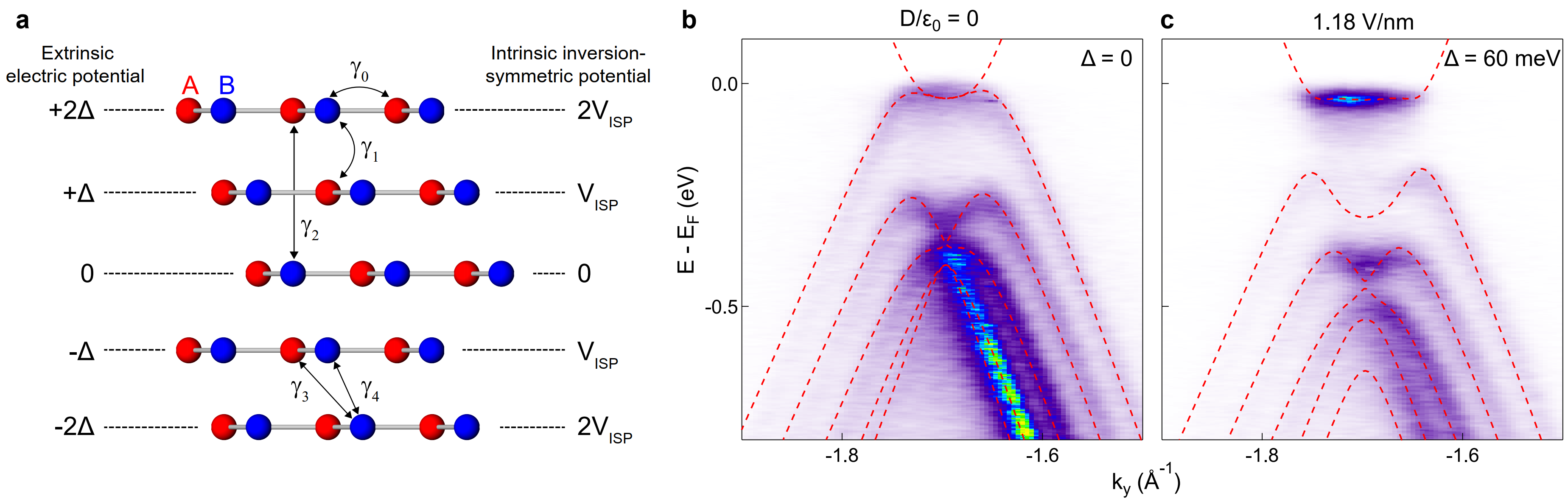}
	\caption{\textbf{Comparison between experimental calculated dispersions.} \textbf{a}, Atomic structure of R5G, where model parameters are labelled. \textbf{b,c}, NanoARPES dispersion measured at $D/\varepsilon_0$ = 0  and 1.18 V/nm, with calculated dispersions for $\Delta$ = 0 and 60 meV appended (red dashed curves).}
	\label{SI_Fig5}
\end{figure*}

\begin{figure*}[htbp]
	\centering
	\includegraphics[width=13cm]{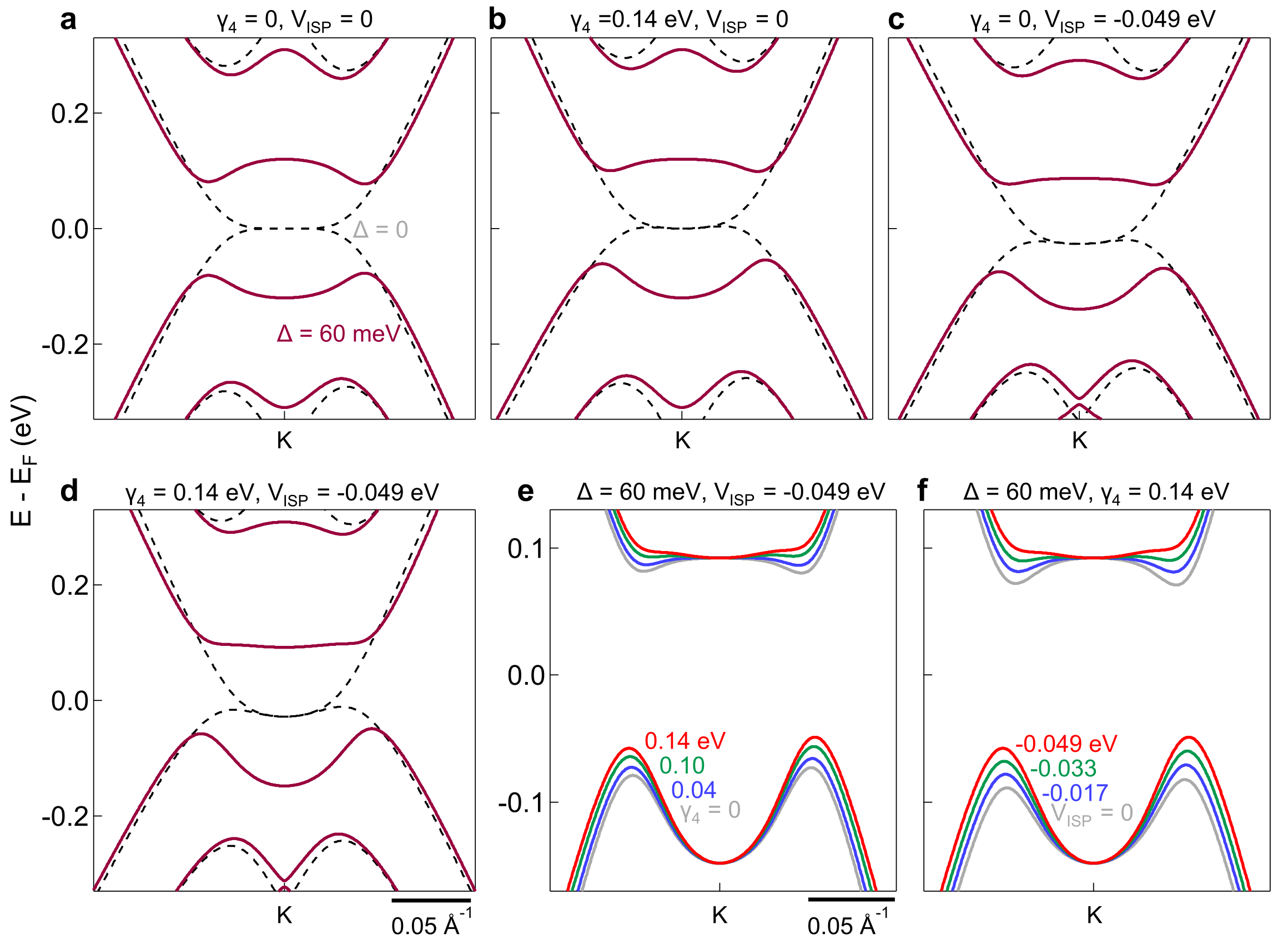}
	\caption{\textbf{Parameter sensitivity of the electron-hole asymmetry.} \textbf{a-d}, Calculated dispersion with and without $\gamma_4$ and $V_{ISP}$. \textbf{e,f}, Calculated dispersion with different choices of $\gamma_4$ and $V_{ISP}$.}
	\label{SI_Fig6}
\end{figure*}

\begin{figure*}[htbp]
	\centering
	\includegraphics[width=16cm]{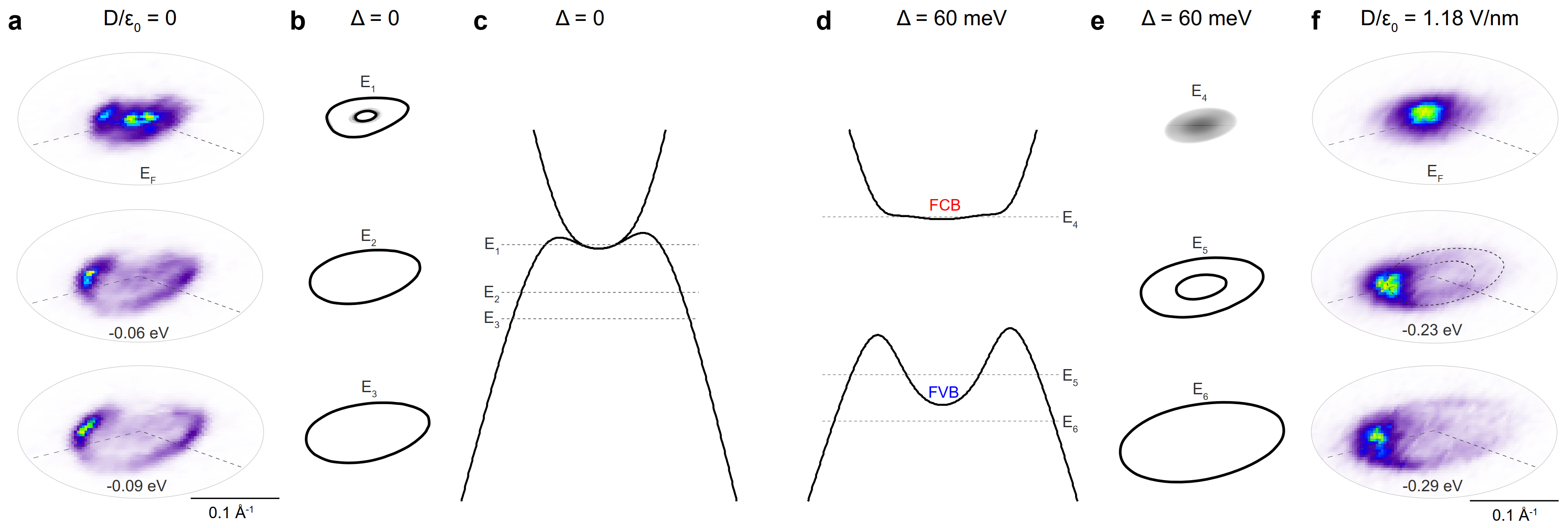}
	\caption{\textbf{Gate-tailored Fermi surface topology (sample S2).} \textbf{a}, NanoARPES energy contours at $E_F$, -0.06 eV and -0.09 eV measured at $D/\varepsilon_0$ = 0 V/nm. \textbf{b}, Corresponding calculated energy contours at $E_1$, $E_2$ and $E_3$ indicated by gray dotted lines in \textbf{c}. \textbf{d}, Calculated dispersion with interlayer potential difference $\Delta$ = 60 meV. \textbf{e}, Calculated energy contours at energies $E_4$, $E_5$ and $E_6$. \textbf{f}, NanoARPES energy contours at $E_F$, -0.23 eV and -0.29 eV measured upon gating at $D/\varepsilon_0$ = 1.18 V/nm.}
	\label{SI_Fig7}
\end{figure*}

\begin{figure*}[htbp]
	\centering
	\includegraphics[width=16cm]{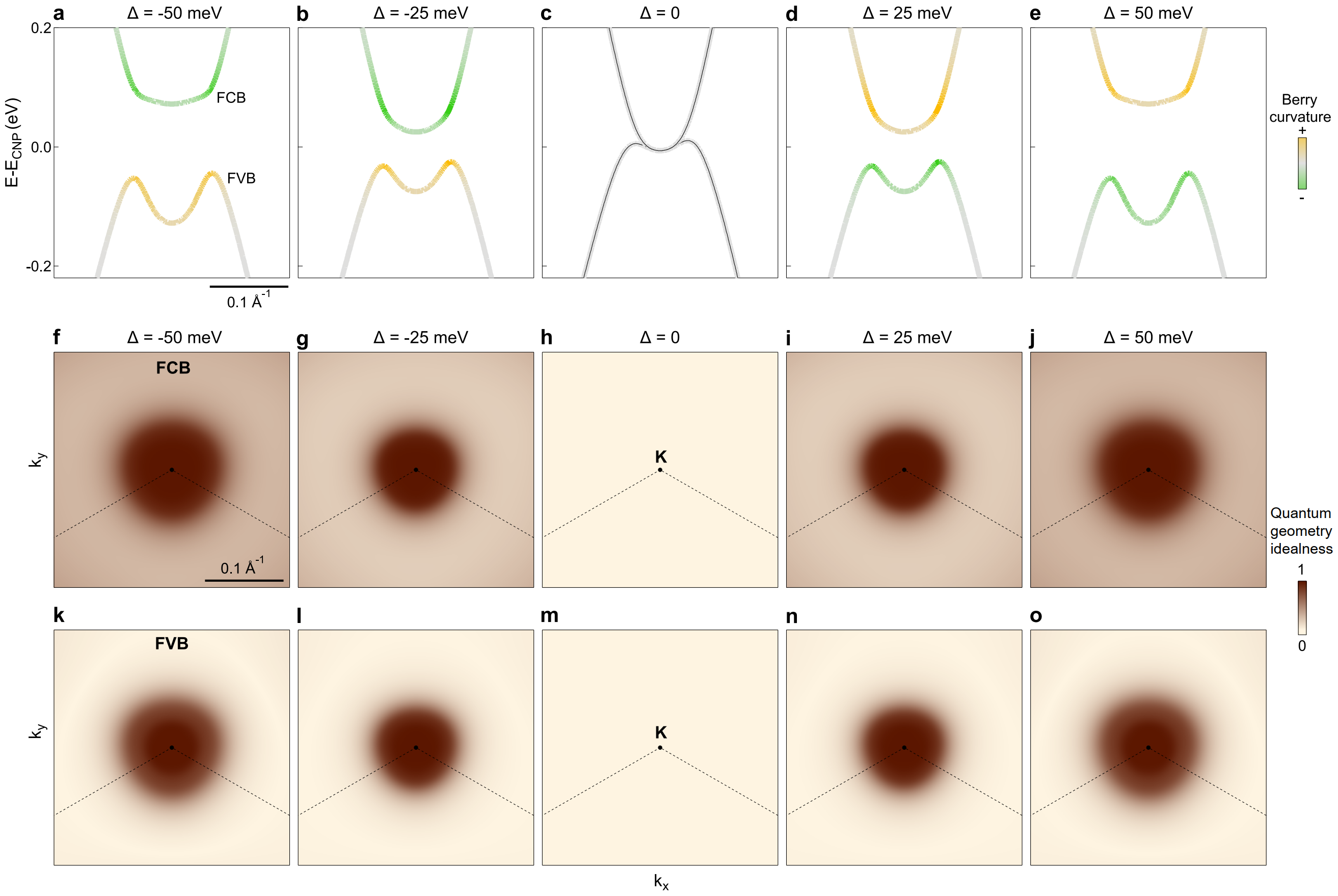}
	\caption{\textbf{Field-induced finite Berry curvature and the ideal quantum geometry.} \textbf{a-e}, Calculated dispersion with the interlayer potential difference of -50, -25, 0, +25, and +50 meV, with the Berry curvature distribution indicated by yellow and green colors. Finite Berry curvature is only observed when applying electric field. \textbf{f-o}, Calculated quantum idealness for FCB (\textbf{f-j}) and FVB (\textbf{k-o}) under different layer potential difference.}
	\label{SI_Fig8}
\end{figure*}

\begin{figure*}[htbp]
	\centering
	\includegraphics[width=16cm]{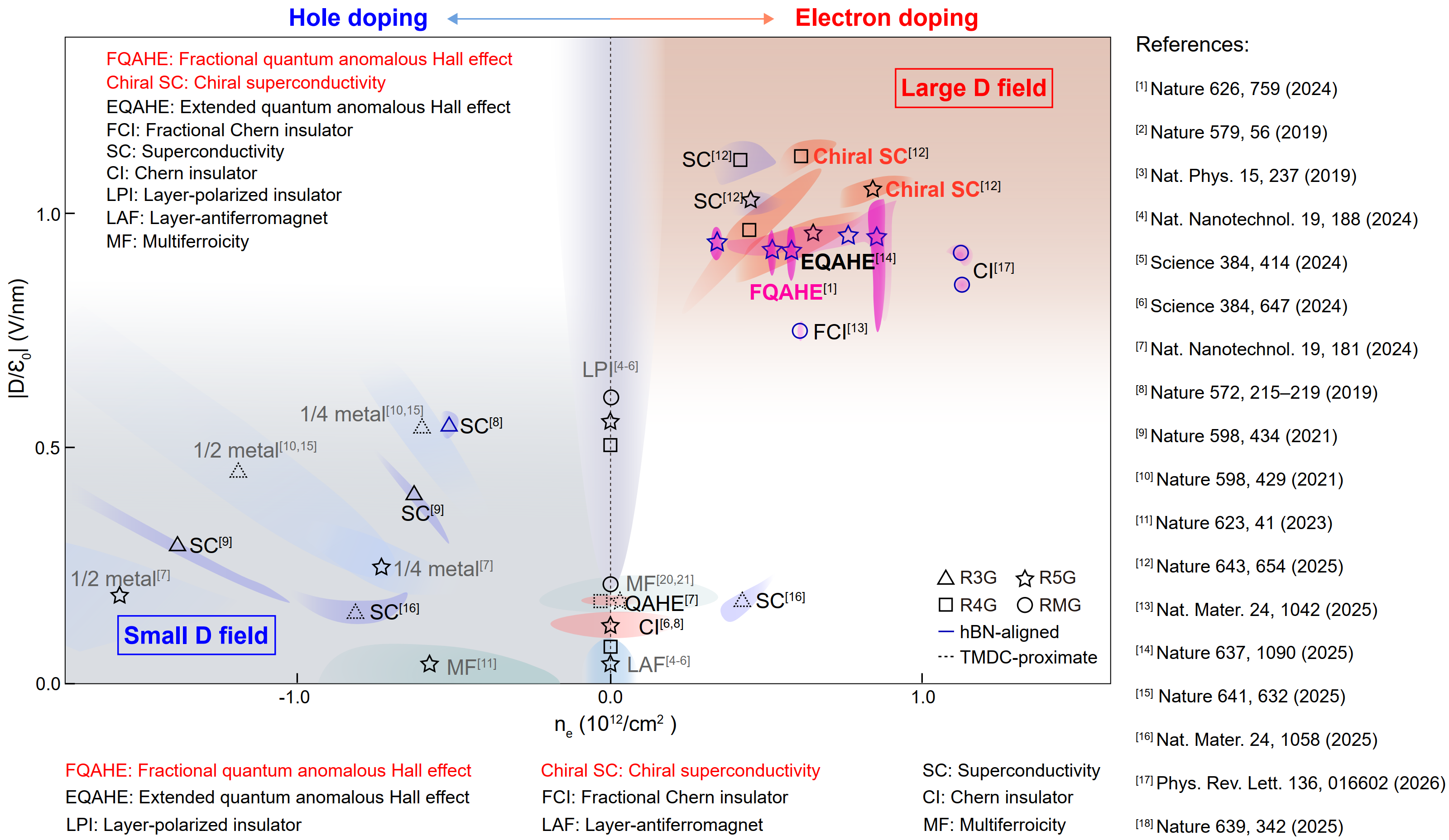}
	\caption{\textbf{Summary of correlated quantum phases in rhombohedral graphene with references added.}}
	\label{SI_Fig9}
\end{figure*}

\end{methods}

\newpage
\newpage
\begin{addendum}
	
	\item[Data availability] The data that support the plots within this paper and other findings of this study are available from the corresponding author upon reasonable request.
	
	\item[Acknowledgement]
	This work is supported by the National Key R $\&$ D Program of China (grant no.~2021YFA1400100), the National Natural Science Foundation of China (grant no.~12522403), Tsinghua University Initiative Scientific Research Program (grant no.~20251080106), National Natural Science Foundation of China (grant no.~12234011, 12421004), and the New Cornerstone Science Foundation through the XPLORER PRIZE. G.C. acknowledges National Natural Science Foundation of China (grant no. 12350005), Shanghai Science and Technology Innovation Action Plan (grant no.~24LZ1401100) and Yangyang Development Fund. Z.S. and Y.W. were supported by National Natural Science Foundation of China (grant no.~12274005), National Key R $\&$ D Program of China (grant no.~2021YFA1401900) and Innovation Program for Quantum Science and Technology (grant no.~2021ZD0302403). T. S. acknowledges support from JST-ASPIRE (grant no. JPMJAP2512). K.L. acknowledges support from the National Natural Science Foundation of China (grant no.~124B2071). K.W. and T.T. acknowledge support from the JSPS KAKENHI (grant no.~21H05233 and 23H02052) , the CREST (JPMJCR24A5), JST and World Premier International Research Center Initiative (WPI), MEXT, Japan. 	
	We acknowledge SOLEIL for the provision of synchrotron radiation facilities of beamline ANTARES, and the Diamond Light Source for the provision of synchrotron radiation facilities of the Beamline I05.	
	
	\item[Author Contributions] S.Z. conceived the research project. H.Z., J.L., F.W., W.C., J.A., P.D., M.D. and S.Z. performed the NanoARPES measurements and analyzed the NanoARPES data. S.W., K.L., H.Z., J.L. and G.C. prepared the samples. S.W., K.L., J.L., L.W., J.Z., P.Y. and G.C. performed the AFM measurements. Y.W. and Z.S. performed the calculations. T.S., P.Y. and W.D. contributed to the discussions. K.W. and T.T. prepared BN crystals. H.Z., J.L., and S.Z. wrote the manuscript, and all authors commented on the manuscript.
	
	\item[Competing Interests] The authors declare that they have no competing financial interests.
	
	\item[Correspondence and requests for materials] should be addressed to Shuyun Zhou (email: syzhou@mail.tsinghua.edu.cn) and  Guorui Chen (chenguorui@sjtu.edu.cn) and Zhida Song (songzd@pku.edu.cn).
	
\end{addendum}

\end{document}